\documentclass[preprint,showkeys, preprintnumbers ,amsmath,amssymb]{revtex4}

\usepackage{graphicx}
\usepackage{dcolumn}
\usepackage{bm}

\begin{document}




\title{ Determination of a wave function functional: The constrained
search variational method}


\author{Xiao-Yin Pan}
\author{ Viraht Sahni}
\author{Lou Massa}
\affiliation{ The Graduate School of the City University of New
York, New York, New York 10016. }

\date{\today}

\begin{abstract}

In a recent paper we proposed the expansion of the space of
variations in energy calculations by considering the approximate
wave function $\psi$ to be a functional of functions $\chi: \psi =
\psi[\chi]$ rather than a function. For the determination of such
a wave function functional, a constrained search is first
performed over the subspace of all functions $\chi$ such that
$\psi[\chi]$ satisfies a physical constraint or leads to the known
value of an observable. A rigorous upper bound to the energy is
then obtained by application of the variational principle. To
demonstrate the advantages of the expansion of variational space,
we apply the constrained-search--variational method to the ground
state of the negative ion of atomic Hydrogen, the Helium atom, and
its isoelectronic sequence. The method is equally applicable to
excited states, and its extension to such states in conjunction
with the theorem of Theophilou is also described.

\end{abstract}


\maketitle

\section*{1. Introduction}

In the traditional application of the variational principle for
the energy \cite{1}, the space of variations is limited by the
choice of the analytical form chosen for the approximate wave
function. For example, if Gaussian or Slater-type orbitals or a
linear combination of such orbitals is employed in the energy
functional of the wave functions, then the variational space is
limited by this choice of functions of the wave functions. In a
recent paper \cite{2} we proposed the idea of overcoming this
limitation by expanding the space over which the variations are
performed. This allows for a greater flexibility for the structure
of the wave function.  A consequence of this greater variational
freedom is that better energies can be obtained.  Or,
equivalently, fewer variational parameters are needed to achieve a
required accuracy.\\

The manner by which the space of variations can be expanded
\emph{in principle} is by considering the approximate wave
function $\psi$ to be a functional of a set of functions $\chi:
\psi= \psi[\chi]$, rather than a function.  The space of
variations is expanded because the functional $\psi[\chi]$ can be
adjusted through the function $\chi$ to reproduce any well behaved
function. The space over which the search for the functions $\chi$
is to be performed, however, is too large for practical purposes,
and a subset of this space must be considered.  The subspace over
which the search for the functions $\chi$ is to be performed is
defined by the requirement that the wave function functional
$\psi[\chi]$ satisfy a constraint. Typical constraints on the
functional $\psi[\chi]$ are those of normalization, the
satisfaction of the Fermi-Coulomb hole sum rule, the requirement
that it lead to observables such as the electron density,
diamagnetic susceptibility, nuclear magnetic constant, Fermi
contact term, or any other physical property of interest.  With
the wave function functional $\psi[\chi]$ thus determined, a
rigorous upper bound to the energy is obtained by application of
the variational principle.  In this manner, a particular property
of interest is obtained \emph{exactly} while simultaneously the
energy is determined accurately.  We refer to this way of
determining an approximate wave function as the
constrained-search--variational method.  The method is general in
that it is applicable to both
ground and excited states.\\

In section $2$ of the paper, we explain the
constrained-search--variational method in further detail.  To
demonstrate the ideas involved, we apply the method in section $3$
to the ground state of the Helium atom, its isoelectronic
sequence, and the negative ion of atomic Hydrogen.  Concluding
remarks are made in section
$4$.\\

\section*{2. Constrained-
search--variational method}

To explain the method for the determination of a wave function
functional, consider the non-relativistic Hamiltonian of the
Helium atom, the ions of its isoelectronic sequence, and the
negative ion of atomic Hydrogen. In atomic units ($e=\hbar=m=1$)
\begin{equation}
\hat{H}=-\frac{1}{2}\nabla_{1}^{2}-\frac{1}{2}\nabla_{2}^{2}-\frac{Z}{r_{1}}
  -\frac{Z}{r_{2}}+\frac{1}{r_{12}},
\end{equation}
where ${\bf r}_{1}$, ${\bf r}_{2}$ are the coordinates of the two
electrons, $r_{12}$ is the distance between them, and $Z$ is the
atomic number. In terms of the Hylleraas coordinates\cite{3}:
$s=r_{1}+r_{2}, \; t=r_{1}-r_{2}, \;  \; u=r_{12}$, which are the
natural coordinates for this atom, we choose the approximate wave
function functional to be of the general form
\begin{equation}
\psi[\chi]=\Phi(s,t,u)[1-f(\chi; s,t,u)],
\end{equation}
 with $\Phi(s,t,u)$ a Slater determinantal pre-factor and $f(\chi;
 s,t,u)]$ a correlated correction term:
 \begin{equation}
f(s,t,u)=e^{-q u}(1+qu)[1-\chi(q;s,t,u)(1+u/2)],
\end{equation}
where $q$ is a variational parameter. Note \emph{any} two electron
wave function in a \textit{ground} or \textit{excited} state maybe
expressed in this  form. The Slater determinant
 may be chosen to be the Hartree-Fock theory wave function \cite{4},
 or  determined self-consistently
within the framework of Quantal Density Functional Theory
\cite{5}. For purposes of explanation, we consider here the
determinant composed of  Hydrogenic functions. Thus, for the
ground state $1^{1}S $ of the Helium atom we have $
\Phi[\alpha,s]=(\alpha^{3}/\pi)e^{-\alpha s}$, and for the excited
triplet $2^{3}S $ state $\Phi[\alpha,s,t]=
\sqrt{2/3}(\alpha^{4}/\pi) e^{-\alpha s} t$. (In the latter, for
explanatory purposes, screening effects are  ignored). Further, we
assume that $\chi$ is a function only of the variable $s$:
$\Psi=\Psi[\chi(q,s)]$.  The approximate wave function  functional
$\Psi[\chi(q,s)]$ for the ground state then satisfies the
electron-electron cusp condition \cite{6}. It also satisfies the
electron-nucleus cusp
condition for $\alpha=Z$.\\

Next consider observables such as the size of the atom,
diamagnetic susceptibility, nuclear magnetic constant, Fermi
contact term, etc, which are  represented by the expectation of
operators $W=r_{1}+r_{2}$, $W=r_{1}^{2}+r_{2}^{2}$,
$W=1/r_{1}+1/r_{2}$, $W=\delta({\bf r}_{1})+\delta({\bf r}_{2})$,
respectively.  For the normalization constraint $W=1$. In terms of
the Hylleraas coordinates, these operators are $W(s)=s$,
$W(s,t)=(s^{2}+t^{2})/2$, $W(s,t)=\frac{4 s}{s^{2}-t^{2}}$,
$W(s,t)=\frac{1}{\pi}[
\frac{\delta(\frac{(s+t)}{2})}{(s+t)^{2}}+\frac{\delta(\frac{(s-t)}{2})}{(s-t)^{2}}]$,
and $W(s,t)=1$. In general, observables can be  represented by
single-particle operators  expressed as $W(s,t)$. The expectation
of the operator $W(s,t)$ is then
\begin{equation}
\langle W \rangle =\int \Psi^{*}[\chi]W \Psi[\chi] d\tau
                  =\langle W_{0} \rangle+\Delta W,
\end{equation}
where (for the ground state)
\begin{equation}
 \langle W \rangle_{0}=\int |\Phi(\alpha,s)|^{2} W(s,t)d\tau,
\end{equation}
\begin{eqnarray}
 \Delta W &=&\int |\Phi(\alpha,s)|^{2} W(s,t)[f^2(q,s,t,u)-2 f(q,s,t,u)]
 d\tau  \\
           &=&2\pi^2 \int_{0}^{\infty} |\Phi(\alpha,s)|^{2} g(s)ds,
\end{eqnarray}
where
\begin{equation}
 g(s)=\int _{0}^{s}udu \int_{0}^{u}dt W(s,t)(s^{2}-t^{2})[f^2(q,s,t,u)-2
 f(q,s,t,u)].
\end{equation}

We now assume that the expectation $ \langle W \rangle$ is known
either through experiment or via some accurate calculation
\cite{7}. As our choice of $\Phi (\alpha; s) $ is analytical, then
both
$ \langle W \rangle_{0}$ and $\Delta W$ are now known. \\

The next step is the constrained search over functions $\chi(q,s)$
for which the expectation $\langle W \rangle$ of Eq.(4) is
obtained. If the parameter $\alpha$ in Eq.(7) is fixed, then there
exist \textit{many} functions $g(s)$ for which the expectation
$\langle W \rangle$ can be obtained. This corresponds to a large
subspace of wave function functionals (See Ref. 2). On the other
hand, if the parameter $\alpha$ is variable, then the only way in
which Eq.(7) can be satisfied is if
\begin{equation}
g(s)=G,
\end{equation}
where $G$ is a determinable constant. \emph{This is equivalent to
the constrained search of all wave function functionals over the
subspace in which Eq.(7) is satisfied.}\\

As an example consider the normalization constraint for which  $
\langle W \rangle= \langle W \rangle_{0}=1$, so that $\Delta W=0$.
Then the only way in which Eq.(7) can be satisfied, (for variable
$\alpha$) is if
\begin{equation}
g(s)=0.
\end{equation}
 This condition is thus equivalent to the constrained search over
 the subspace of all normalized functionals $\Psi[\chi(q,s)]$.\\

 Substitution of $f(\chi; s,t,u)$ into Eq.(10) leads to
 a quadratic equation for the function
$\chi(q,s)$:
\begin{equation}
 a(q,s)\chi(q,s)^{2}+2 b(q,s) \chi(q,s)+c(q,s)=0,
\end{equation}
where

\begin{equation}
 a(q,s)=\int_{0}^{s} (s^{2}u^{2}-u^{4}/3)(1+u/2)^{2}(1+qu)^{2}e^{-2 qu}du,
\end{equation}
\begin{eqnarray}
 b(q,s)&=&-\int_{0}^{s}(s^{2}u^{2}-u^{4}/3)(1+u/2)(1+qu) \nonumber\\
     & &[e^{-2 qu}(1+qu)-e^{-qu}]du,
\end{eqnarray}
\begin{equation}
 c(q,s)=\int_{0}^{s} (s^{2}u^{2}-u^{4}/3)(1+qu)[e^{-2 qu}(1+qu)-2
 e^{-qu}]du.
\end{equation}
The integrals for the coefficents   $a(q,s)$, $b(q,s)$, and
$c(q,s)$ are  determined analytically. Solution of the quadratic
equation is equivalent to searching over the entire subspace of
normalized wave function functionals. In this example, the
subspace corresponds to two points. The two solutions
$\chi_{1}(q;s)$ and $\chi_{2}(q;s)$ lead to two normalized wave
functions $\psi[\chi_{1}]$ and $\psi[\chi_{2}]$.\\

The generalization to the case when $W=W(s)$ or $W=W(s,t)$ follows
readily. In either case, one has also to solve a quadratic
equation for the determination of the functions $\chi(q, \alpha;
s)$. One thus obtains two  wave
function functionals that lead to the exact value for $\langle W \rangle$.\\

  For the normalized wave function functionals determined above,
  the energy functional  in terms of $(s,t,u)$ coordinates which is
\begin{eqnarray}
I[\psi[\chi]] &=& \int\psi^{*}{\hat H}\psi d\tau  \\
&=& 2 \pi^{2}
\int_{0}^{\infty}ds \int_{0}^{s}du \int_{0}^{u} dt \{u
(s^{2}-t^{2}) [ (\frac{\partial \psi}{\partial s})^{2}+
(\frac{\partial \psi}{\partial t})^{2}+(\frac{\partial
\psi}{\partial
u})^{2}]\nonumber \\
& & +2 \frac{\partial \psi }{\partial u}[s (u^{2}-t^{2})
\frac{\partial \psi}{\partial s}+ t (s^{2}-u^{2}) \frac{\partial
\psi}{ \partial
t}] \nonumber \\
&&-[4 Z s u-(s^{2}-t^{2})]\psi^{2}\},
\end{eqnarray}
is then minimized with respect to the parameters $\alpha$ and $q$.
(The prefactor minimizes the energy at $\alpha=Z-5/16$). \\

For wave function functionals determined by sum rules other than
normalization, the functional $I[\psi[\chi]]$ must be divided by
the normalization integral $\int \psi^{*}\psi d\tau $. In this
manner, the wave function functionals $\psi[\chi]$ are normalized
, obtain the exact value of the expectation $\langle
W(s,t)\rangle$, and lead to an accurate value for the ground state
energy.
\\

\section*{3. Application to two-electron atomic and ionic systems}

In this section, we apply the constrained-search—-variational
method to two-electron atomic and ionic systems.  The two wave
function functionals $\psi[\chi_{1}]$ and $\psi[\chi_{2}]$
employed are those determined via the constraint of normalization
as described in the previous section with the crude Hydrogenic
prefactor.  In Table I we quote the values for the ground state
energy for $H^{-}$, the He atom, and its isoelectronic sequence.
For the He atom we also quote the values of Hartree-Fock theory
\cite{4}, the $3$-parameter Caratzoulas-Knowles wave function
\cite{8}, and the $1078$-parameter Pekeris wave function \cite{7}.
For $H^{-}$ and the other negative ions corresponding to $Z =
3-8$, we give the values of the variational-perturbation results
of Aashamar \cite{9}.  The satisfaction of the virial theorem and
the percent errors as compared to the Pekeris and Aashamar values
are also given.  The functions $\chi_{1}(q,s)$ and $\chi_{2}(q,s)$
for $H^{-}$, $B^{3+}$, and $O^{6+}$ are plotted in
Figs. 1-3. \\

\begin {table}

\caption{\label{Table 1.} Rigorous upper bounds to the ground
state of $H^{-}$, $He$, $Li^{+}$, $Be^{2+}$, $B^{3+}$, $C^{4+}$,
$N^{5+}$, $O^{6+}$, in atomic units as determined from the wave
function functionals determined via the constraint of
normalization together with the values due to Hartree-Fock (HF)
theory \cite{4}, Caratzoulas-Knowles(CK)\cite{8}, Pekeris\cite{7}
and Aashamar\cite{9}. The satisfaction of the virial theorem, and
the percent errors compared to the values of Pekeris and Aashamar
are also given. } \vspace{10mm}

\renewcommand{\arraystretch}{0.4}
\begin{tabular}{|c |c |c | c|c | c |}
\hline \hline  Ion or Atom & Wave function & Parameters &
Ground state energy & $\%$ error & $-V/T$  \\
\hline

\large{$H^{-}$} & \large{$\Phi$} & &\large{$-0.473$} &
  \large{$10.37$}  & \large{ $2.0000$} \\   \cline{2-6}

         &\large{$\psi[\chi_{1}]$}&  $\alpha=0.6757$, $q=0$ &\large{$-0.50946$}  &\large{$3.486$}&
         \large{$2.0019$} \\ \cline{2-6}

         &\large{$\psi[\chi_{2}]$}& $\alpha=0.6757$, $q=0$ &\large{$-0.50946$}  &\large{$3.486$}& \large{$2.0019$}  \\ \cline{2-6}
       & \large{Aashamar}  &   & \large{$-0.52775 $}  &   & \large{ $2.0000$}     \\
       \hline \hline

\large{$He$} & \large{$\Phi$} & & \large{$-2.84766$} &
\large{$1.931$} & \large{ $2.0000$} \\   \cline{2-6}

         &\large{$\psi[\chi_{1}]$}& $\alpha=1.6614, q=0.5333$ &\large{$-2.89072$}  &\large{$0.448$}&
         \large{$1.9973$} \\ \cline{2-6}

         &\large{$\psi[\chi_{2}]$}& $\alpha=1.6629, q=0.1705$  &\large{$-2.89122$}  &\large{$0.430$}& \large{$1.9984$}  \\ \cline{2-6}
& \large{HF}   &  & \large{$-2.86168 $}  & \large{$1.448$}  &
\large{ $2.0000$}     \\\cline{2-6}
 & \large{CK}   &  & \large{$-2.89007 $}  & \large{$0.470$}  & \large{ $1.9890$}     \\\cline{2-6}
       & \large{Pekeris}   &  & \large{$-2.90372 $}  &    & \large{ $2.0000$}     \\
       \hline \hline

\large{$Li^{+}$} & \large{$\Phi$} &  &\large{$-7.22266$} &
\large{$0.786$}  & \large{ $2.0000$} \\   \cline{2-6}

         &\large{$\psi[\chi_{1}]$}&$\alpha=2.6595, q=1.2287$  &\large{$-7.26687$}  &\large{$0.179$}&
         \large{$1.9981$} \\ \cline{2-6}

         &\large{$\psi[\chi_{2}]$}& $\alpha=2.6610, q=0.2897$ &\large{$-7.26820$}  &\large{$0.161$}& \large{$1.9992$}  \\ \cline{2-6}
       & \large{Aashamar}   &  & \large{$-7.27991 $}  &   & \large{ $2.0000$}     \\
       \hline \hline

\large{$Be^{2+}$} &  \large{$\Phi$} &  &\large{$-13.59766$} &
\large{$0.424$} & \large{ $2.0000$} \\   \cline{2-6}

         &\large{$\psi[\chi_{1}]$}& $\alpha=3.6584, q=1.8950$ & \large{$-13.64219$}  &\large{$0.098$}&
         \large{$1.9987$} \\ \cline{2-6}

         &\large{$\psi[\chi_{2}]$}& $\alpha=3.6599, q=0.3722$ &\large{$-13.64416$}  &\large{$0.084$}& \large{$1.9995$}  \\ \cline{2-6}
       & \large{Aashamar}     & &\large{$-13.65557 $}  &   & \large{ $2.0000$}     \\
       \hline \hline

\large{$B^{3+}$} & \large{$\Phi$} & &\large{$-21.97266$} &
\large{$0.265$} & \large{ $2.0000$} \\   \cline{2-6}

         &\large{$\psi[\chi_{1}]$}& $\alpha=4.6578, q=2.5711$ &\large{$-22.01729$}  &\large{$0.062$}&
         \large{$1.9991$} \\ \cline{2-6}

         &\large{$\psi[\chi_{2}]$}& $\alpha=4.6592, q=0.4401$ &\large{$-22.01973$}  &\large{$0.051$}& \large{$1.9997$}  \\ \cline{2-6}
       & \large{Aashamar}     &  &\large{$-22.03097 $}  &   & \large{ $2.0000$}     \\
       \hline \hline

\large{$C^{4+}$} & \large{$\Phi$} & &\large{$-32.34766$} &
\large{$0.181$} & \large{ $2.0000$} \\   \cline{2-6}

         &\large{$\psi[\chi_{1}]$}& $\alpha=5.6574, q=3.2528$ &\large{$-32.39230$}  &\large{$0.043$}&
         \large{$1.9993$} \\ \cline{2-6}

         &\large{$\psi[\chi_{2}]$}& $\alpha=5.6578, q=0.4839$&\large{$-32.39511$}  &\large{$0.034$}& \large{$1.9997$}  \\ \cline{2-6}
       & \large{Aashamar}     & &\large{$-32.40625 $}  &   & \large{ $2.0000$}     \\
       \hline \hline

\large{$N^{5+}$} & \large{$\Phi$}& & \large{$-44.72266$} &
\large{$0.131$} & \large{ $2.0000$} \\   \cline{2-6}

         &\large{$\psi[\chi_{1}]$}&$\alpha=6.6572, q=3.9381$ &\large{$-44.76729$}  &\large{$0.032$}&
         \large{$1.9995$} \\ \cline{2-6}

         &\large{$\psi[\chi_{2}]$}&$\alpha=6.6584, q=0.5511$ &\large{$-44.77035$}  &\large{$0.025$}& \large{$1.9998$}  \\ \cline{2-6}
       & \large{Aashamar}   &  & \large{$-44.78145 $}  &   & \large{ $2.0000$}     \\
       \hline \hline

\large{$O^{6+}$} & \large{$\Phi$} & &\large{$-59.09766$} &
\large{$0.100$} & \large{ $2.0000$} \\   \cline{2-6}

         &\large{$\psi[\chi_{1}]$}&$\alpha=7.6570, q=4.6257$ &\large{$-59.14226$}  &\large{$0.024$}&
         \large{$1.9996$} \\ \cline{2-6}

         &\large{$\psi[\chi_{2}]$}& $\alpha=7.6582, q=0.5985$ &\large{$-59.14554$}  &\large{$0.019$}& \large{$1.9998$}  \\ \cline{2-6}
       & \large{Aashamar}   &  & \large{$-59.15660 $}  &   & \large{ $2.0000$}     \\
       \hline \hline

\end{tabular}
\end{table}



\begin{figure}
 \begin{center}
 \includegraphics[bb=0 0 502 665, angle=90, scale=0.7]{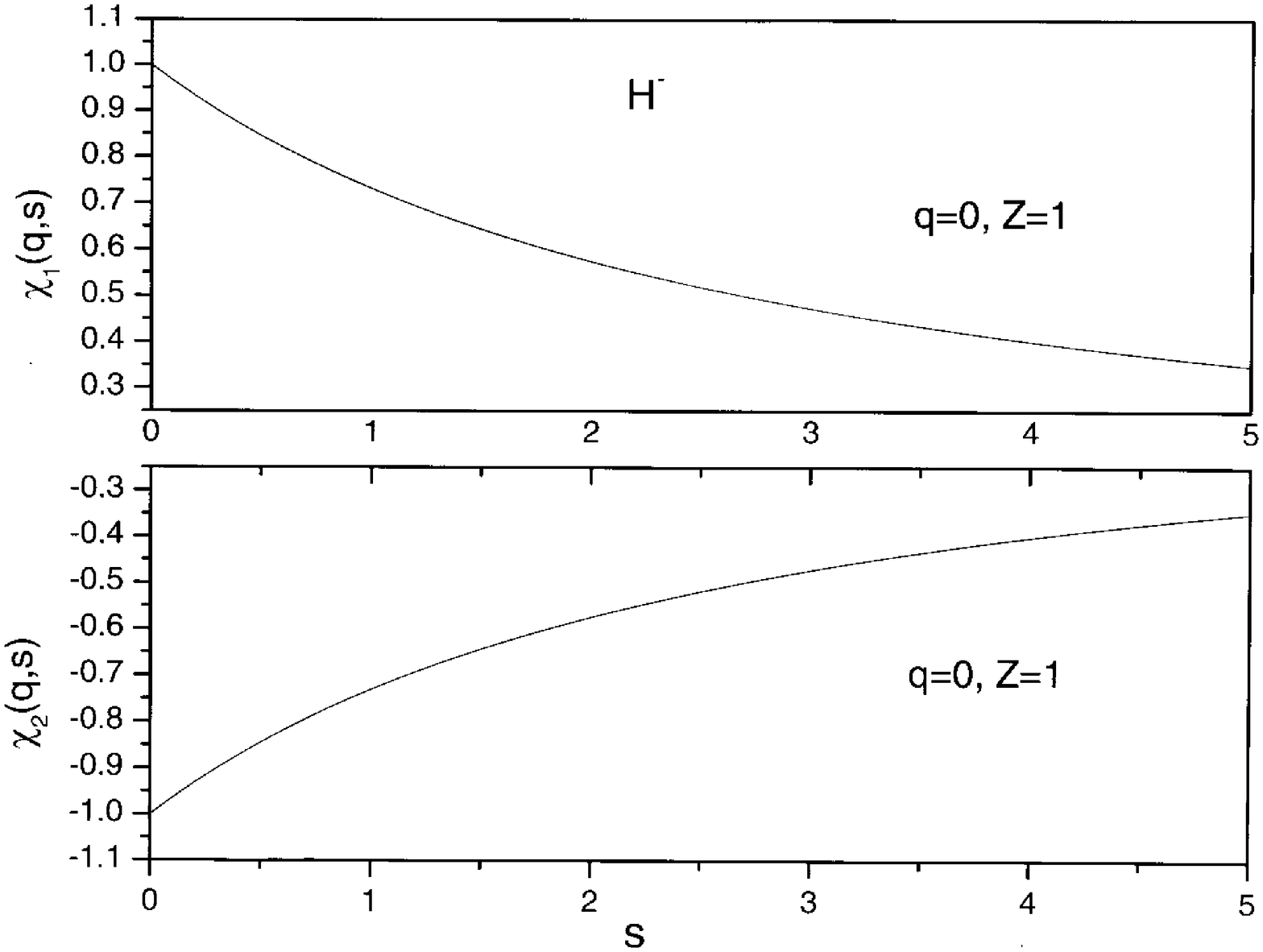}
 \caption{Fig.1:  The functions $\chi_{1}(q,s)$ and $\chi_{2}(q,s)$ for
$H^{-}$. \label{}}
 \end{center}
 \end{figure}


\begin{figure}
 \begin{center}
 \includegraphics[bb=0 0 502 665, angle=269.5, scale=0.7]{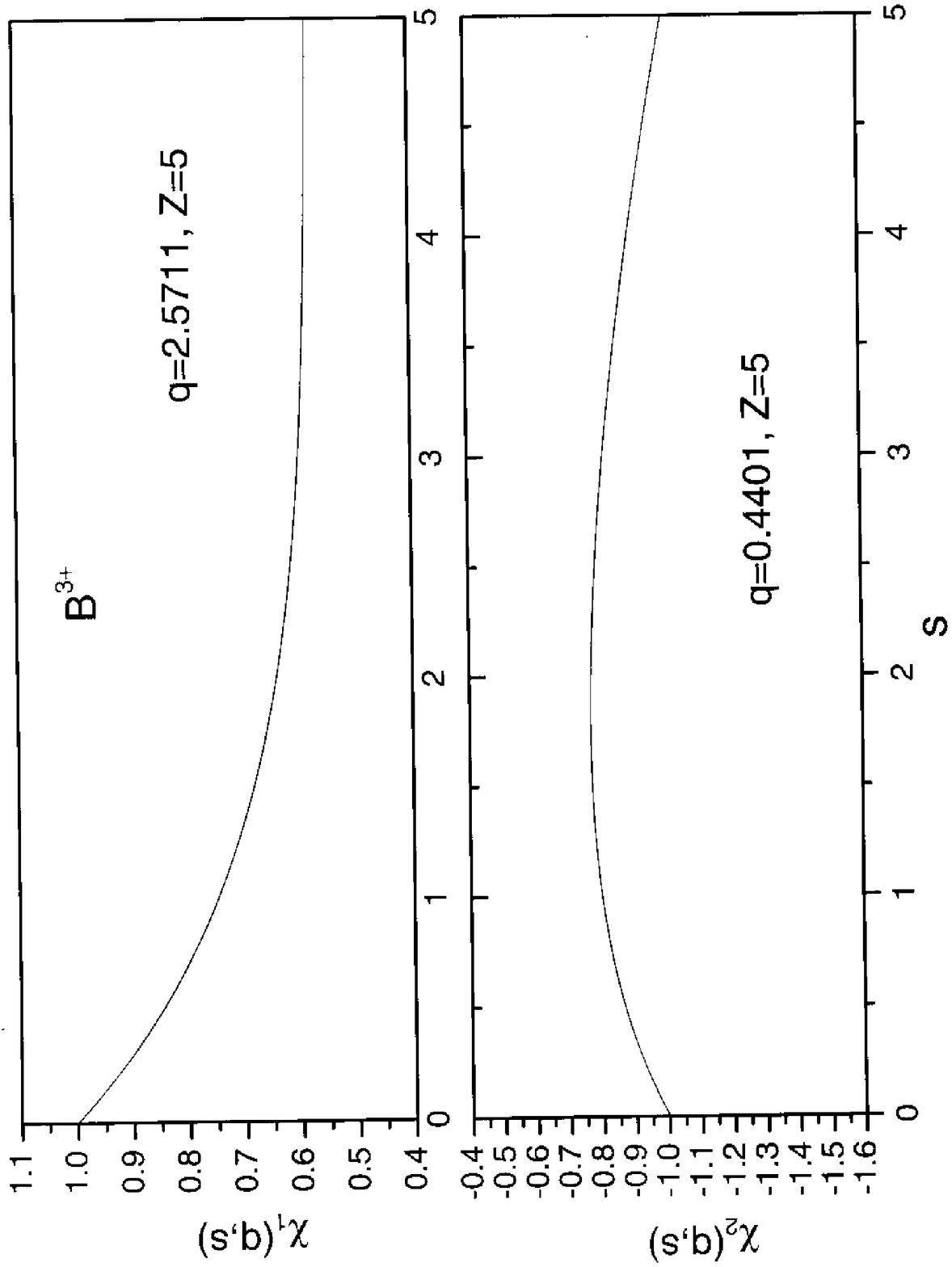}
 \caption{Fig.2:  The functions $\chi_{1}(q,s)$ and $\chi_{2}(q,s)$ for
$B^{3+}$.\label{}}
 \end{center}
 \end{figure}


\begin{figure}
 \begin{center}
 \includegraphics[bb=0 0 502 665, angle=270, scale=0.7]{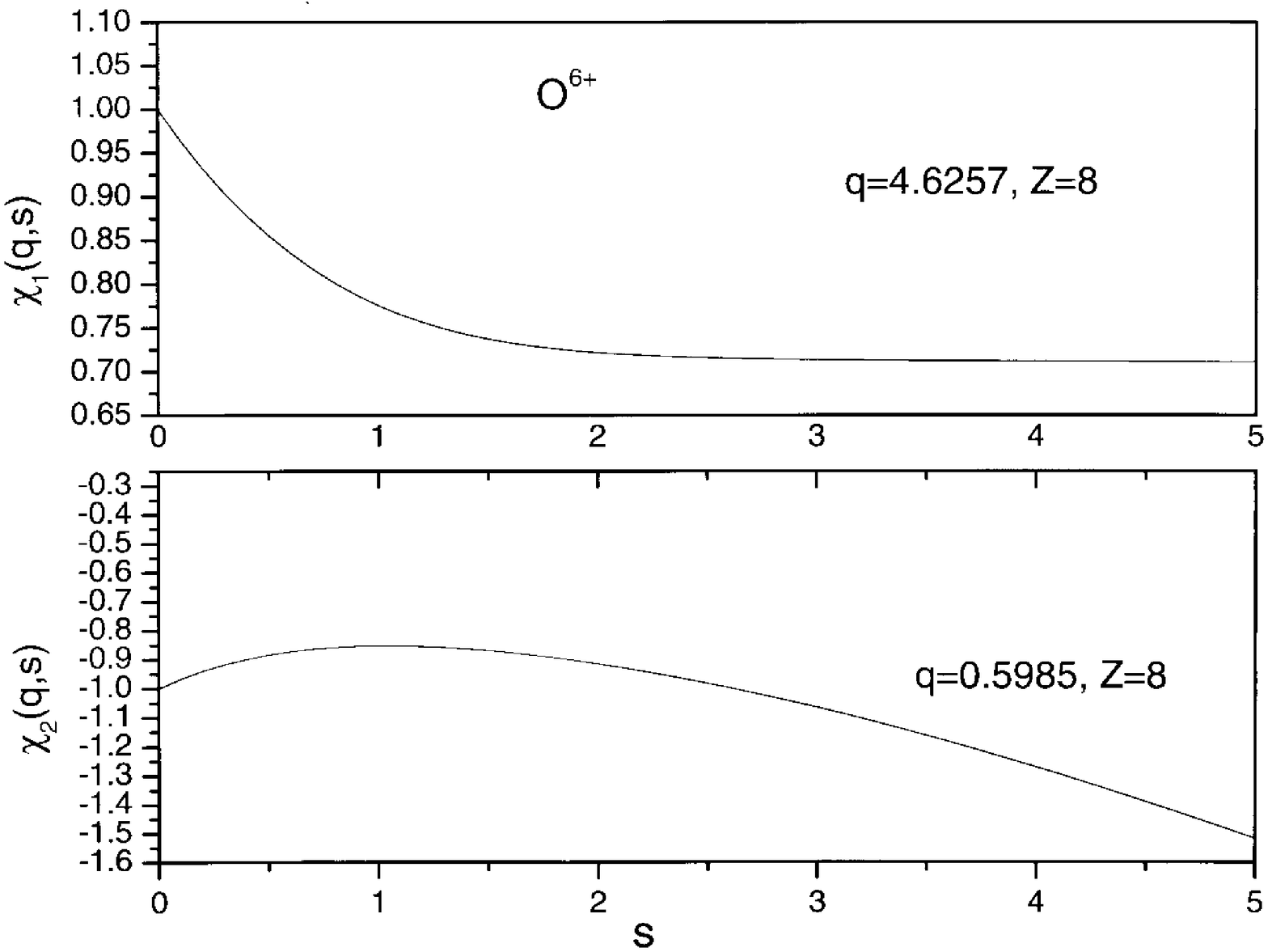}
 \caption{
Fig.3: The functions $\chi_{1}(q,s)$ and $\chi_{2}(q,s)$ for
$O^{6+}$.\label{}}
 \end{center}
 \end{figure}

Observe that the improvement of the energies of the two wave
function functionals over the prefactor values is generally an
order of magnitude.  As expected, the energies as well as the
satisfaction of the virial theorem improves with increasing atomic
number $Z$.  For the He atom, the energies of both
$\psi[\chi_{1}]$ and $\psi[\chi_{2}]$ are superior to those of
Hartree-Fock theory and of the 3-parameter Caratzoulas-Knowles
wave function. Furthermore, whereas the prefactor leads to a
negative electron affinity, both wave function functionals lead to
a positive electron affinity for $H^{-}$  as must be the case as
$H^{-}$  is stable.  The exact satisfaction of the virial theorem
by the prefactor is a consequence of scaling, whereas that of
Hartree-Fock theory is
because of self-consistency.\\

In Table II we quote the values of the operators $W
=\sum_{i=1}^{2} r^{n}, n = -2, -1, 1, 2$, and $W = \delta({\bf
r}_{1}) +\delta({\bf r}_{2})$ for the He atom as determined by
both $\psi[\chi_{1}]$ and $\psi[\chi_{2}]$ together with the
Hartree-Fock theory, Caratzoulas-Knowles, and Pekeris values. The
accuracy of these results is, of course, not correct to second
order as are those for the energy. Nonetheless, the results are
considerable improvements over the prefactor values. They are also
all superior to the 3-parameter results of Caratzoulas-Knowles.
The latter indicates that the two wave function functionals
although also determined via energy minimization, are superior
throughout space. Thus, by expanding the space of variations, one
obtains a superior wave function not only in the region
contributing most to the energy, but also in other regions of
space.  The superiority of the Hartree-Fock theory values, on the
other hand, is due to the fact that in this theory, the
expectations of single-particle
operators is correct to second order \cite{10}.\\

\begin{table}
\caption{\label{ } The expectation value of the operator
$W=\sum_{i=1}^{2}r_{i}^{n}; n=-2,-1,1,2 $ and $W=\delta({\bf
r}_{1})+\delta({\bf r}_{2})$ for the $He$ atom employing the wave
function functionals determined by the normalization constraint,
and by the Hartree-Fock theory(HF)\cite{4},
Caratzoulas-Knowles(CK)\cite{8}, and Pekeris \cite{7}  wave
functions (WF). } \vspace{10mm}
\renewcommand{\arraystretch}{1.0}
\begin{tabular}{|c |c |c|c|c|c|}
\hline \hline WF  &$<(1/r_{1}+1/r_{2})>$&
$<(1/r_{1}^{2}+1/r_{2}^{2})>$& $<(r_{1}^{2}+r_{2}^{2})>$ &
$<r_{1}+r_{2}>$ & $\langle\delta({\bf r}_{1})+\delta({\bf
r}_{2})\rangle$
\\ \hline
$\Phi$ &  $3.3750$ & $11.391$ & $2.1069$ & $1.7778$ & $3.05922$\\
\hline

 $\psi[\chi_{1}]$& $3.3773$& $11.726$& $2.1924$& $1.8057$ & $3.37921$\\\hline
 $\psi[\chi_{2}]$& $3.3784$& $11.727$ & $2.1876$ & $1.8041$ &$3.37925$ \\\hline \hline
 $HF$ &  $3.3746$ & $11.991$& $2.3697$ & $1.8545$& $3.5964$\\ \hline
$CK$ & $3.3911$ & $11.714$ & $2.1292$ & $1.7848$ &
\\ \hline
$Pekeris$ & $3.3766$&$12.035$&$2.3870$&$1.8589$& $3.6208$\\
\hline \hline
\end{tabular}
\end{table}

We note that the two functions $\chi_{1}(q,s)$ and $\chi_{2}(q,s)$
are very different from each other: $\chi_{1}(q,s)$ is positive
whereas $\chi_{2}(q,s)$ is negative.  See Figs. 1-3.  Thus,
although the analytical form of the wave function is the same, the
two functionals $\psi[\chi_{1}]$ and $\psi[\chi_{2}]$ are very
different. Nevertheless, they lead to accurate results that are
essentially the same. Thus, the constrained search  for the
functions $\chi$ over this subspace of normalized wave functions
leads to two physically meaningful functionals.\\

 \section*{4. Concluding remarks}

  In this paper we have shown how to expand the space of variations
in calculations of the energy by constructing approximate wave
functions that are functionals rather than functions.  The wave
function functionals depend upon functions that are chosen so as
to satisfy a sum rule or reproduce the value of an observable.  In
this constrained-search—-variational method, wave functions that
are accurate over all space are thereby obtained.  The framework
presented is general and applicable to both ground and excited
states.  For excited states, one would in addition employ the
theorem of Theophilou \cite{11} according to which if
$\varphi_{1}, \varphi_{2},…,\varphi_{m}$,..., are orthonormal
trial functions for the $m $ lowest eigenstates of the Hamiltonian
$H$, having exact eigenvalues $E_{1}, E_{2}, …E_{m}$,... , then
$\sum_{i=1}^{m} \langle \varphi_{i} |H |\varphi_{i}\rangle \geq
\sum_{i=1}^{m} E_{i}$ . In this way, a rigorous upper bound to the
\emph{sum }of the ground and excited states is achieved. With the
ground state energy known, a rigorous upper bound to the excited
state energy is then determined, while simultaneously a physical
constraint or sum rule is satisfied or an observable
obtained exactly.\\

 In the calculations presented to demonstrate these ideas, a crude Hydrogenic Slater
 determinantal prefactor was employed.  Improved results may be obtained through a
  better prefactor.  Fully self-consistently determined prefactors for many-electron
  systems may be achieved, for example, via Quantal density functional theory.
  The latter is a local effective potential energy theory of noninteracting Fermions
   with the true density in which the multiplicative potential energy operator representative
    of all the many-body effects is explicitly defined in terms of the interacting system
     wave function and the orbitals of this model system.  These orbitals, determined
      self-consistently, then constitute the Slater determinantal prefactor.
      Or one could employ analytical or self-consistently determined Hartree-Fock theory
       orbitals for the prefactor.  Another step towards improved results would be to further
        expand the space of variations defining the functions $\chi$.  In such a case, the equation
         for the functions $\chi$ could be an integral equation.  Other analytical forms for the
         correlation factor could also be employed.  These avenues are being pursued
         to better understand the ideas underlying the construction of wave function functionals,
          and to employ these functionals within the context of Quantal density functional
          theory.\\

\begin{acknowledgments}
This work was supported in part by the Research Foundation of
 CUNY. L. M. was supported in part by NSF through CREST, and by
 a ``Research Centers in Minority Institutions'' award, RR-03037,
 from the National Center for Research Resources, National
 Institutes of Health. \\
\end{acknowledgments}






\begin{references}
\bibitem{1}
B. L. Moiseiwitsch, \emph{Variational Principles}, John Wiley and
Sons, (New York , 1966).

\bibitem{2}X.-Y. Pan, V. Sahni, and L. Massa, Phys. Rev.
Lett. \textbf{93}, 130401 (2004).

\bibitem{3} E. A. Hylleraas, Z. Physik, \textbf{48}, 469
(1928); X.-Y Pan, V. Sahni,  and L. Massa, \emph{physics/0310128}.


\bibitem{4} C. F. Fischer, \emph{The Hartree-Fock Method for Atoms}, John Wiley
and Sons, (New York, 1977).

\bibitem{5}V. Sahni,
\emph{Quantal  Density Functional Theory}, Springer-Verlag,
(Berlin, 2004).

\bibitem{6}X.-Y. Pan and V. Sahni, J. Chem. Phys. \textbf{119}, 7083 (2003); R.
T. Pack and W. Byers Brown, J. Chem. Phys. \textbf{45}, 556
b(1966); W. A. Bingel, Theor. Chim. Acta \textbf{8}, 54 (1967).

\bibitem{7}C. L. Pekeris, Phys. Rev. \textbf{115}, 1216 (1959).

\bibitem{8}S. Caratzoulas and P. J. Knowles, Mol. Phys. \textbf{98},
1811 (2000).

\bibitem{9} K. Aashamar, Physica Mathematica, University of
Osloensis, Report No. \textbf{35} and \textbf{36} (1969).

\bibitem{10} J. Goodisman and Klemperer, J. Chem. Phys. \textbf{38}, 721 (1963).

\bibitem{11} A. Theophilou, J. Phys.
C \textbf{12}, 5419 (1979).
\end{references}
\end{document}